# Virtual surfaces, director domains and the Fréedericksz transition in polymer stabilized nematic liquid crystals

Pavel A. Kossyrev, Jun Qi, Nikolai V. Priezjev, Robert A. Pelcovits, and Gregory P. Crawford

Department of Physics, Brown University, Providence, Rhode Island 02912, USA

**Abstract**

The critical field of the Fréedericksz transition and switching dynamics are investigated for Polymer Stabilized Nematic Liquid Crystals as a function of polymer concentration. A simple phenomenological model is proposed to describe the observed critical field and dynamic response time behaviors as a function of concentration. In this model the polymer fibrils form director domains, which are bounded by "virtual surfaces" with a finite anchoring energy. The Fréedericksz transition occurs independently within each of these domains.

Polymer Stabilized Liquid Crystals (PSLC) are composed of a low concentration polymer network dispersed in a nematic liquid crystal. They are promising materials for many display configurations. Small amounts of reactive monomer polymerized in a nematic liquid crystal create an aligned polymer network that under some circumstances can provide a memory effect.[1] The polymer network structure captures the details of the nematic director prior to photo-polymerization. In this paper we present results on the influence of the polymer network on the switching behavior of a homogeneously aligned nematic liquid crystal. We show that the critical field of the Fréedericksz transition and the dynamic response time are strongly influenced by the presence of the polymer network.

Previous studies calculated the critical field of the Fréedericksz transition in PSLC systems by modeling the polymer morphology as collection of planes or bundles parallel to the substrates,[2,3] or collectively as an effective field.[4] The model of the polymer matrix as a series of planes perpendicular to the $x$-, $y$-, and $z$-axes of the Cartesian coordinate system was introduced by Hikmet.[5] None of these models explicitly account for the observed dependence of the critical field on polymer concentration. In this paper we introduce a phenomenological model, which successfully describes the dependence of the critical field and the response time on the polymer concentration, and accounts for the change in the polymer network's characteristic length and anchoring strength as the polymer concentration is varied.

Consider a planar cell of PSLC sandwiched between two transparent indium tin oxide (ITO) coated glass plates with a 5 μm cell gap thickness. The glass substrates are coated with rubbed polyimide (Nissan-0821) such that the nematic directors on both surfaces are anchored parallel to the plates. We prepared two sets of the PSLC using the low molecular weight nematic liquid crystal BL038 (EM Industries) and diacrylate monomers: LC242 (BASF) and RM257 (EM Industries). A low percentage (2%) of photo-initiator Darocur-1173 (Ciba) was used so that polymerization can be carried out in the UV. In our studies we investigated the concentration range where the diacrylate monomers are totally dissolved in liquid crystal at room temperatures (≤6 % of monomer by weight). When a diacrylate monomer is polymerized in a liquid crystal solvent, the orientation and order of the resultant network depends on the orientation and order of the liquid crystal.[1] Thus, we assume that the polymer network has planar alignment identical to the alignment of the liquid crystal.[3,4]

A schematic illustration of the experimental set up for measuring the transmission properties of a PSLC cell is shown in Figure 1(a). The cell is situated between a pair of parallel polarizers, with the nematic director oriented at 45° to the transmission axes of the polarizers. Glass plates coated with ITO serve as the electrodes to apply 1 kHz square ac-voltage, $V$. In the field-off state, the cell exhibits a phase retardation $\Gamma=2\pi/\lambda(n_e-n_o)d\approx 4\pi$, where $n_e=1.799$, $n_o=1.527$ are the extra-ordinary and ordinary refractive indices of the BL038 at 589 nm respectively, $d=5$ μm is the cell gap thickness, and $\lambda=633$ nm is the wavelength of the incident laser light. Due to the uniform alignment of the liquid crystal and the polymer network, the transmission light intensity of the cell $I=1/2\cos^2(\Gamma/2)$ is maximed in the field-off state with the present arrangement of polarizers. A transparent cell can be switched to an opaque state when an electric field is applied, due to the creation of director domains in the cell.[6] By application of the field only a portion of the molecules reorient. Those molecules, which are in close vicinity to the polymer network, remain less influenced by the field due to anchoring to the polymer. In this way domains with different orientations of the nematic director can be created. The phase retardation of the reoriented molecules becomes $\Gamma = 2\pi/\lambda \int (n_e(\theta) - n_o) dx$, where the integral is over the domain size and $n_e(\theta)$ is given by:

$$\left(\frac{1}{n_e(\theta)}\right)^2 = \left(\frac{\sin\theta}{n_o}\right)^2 + \left(\frac{\cos\theta}{n_e}\right)^2 \tag{1}$$

The angle $\theta = \theta(x,y)$ is the tilt angle of the director with respect to the $z$-axis (see Figure 1(a)), which depends on the position $x$, $y$ of the molecule in the domain and the field strength. The transmittance $T = I(V)/I(V=0)$ of the cell in the field-on state is affected by two factors: scattering by the nematic director domains and the change in the phase retardation. The scattering leads to the decrease in the transmittance and to the opaqueness of the PSLC cells. The change in the phase retardation gives rise to oscillations in the transmittance. An example of the transmittance versus applied field is shown in Figure 1(b).

By monitoring the transmittance of the cells as a function of the applied voltage we determine the critical voltage $V_c$ of the Fréedericksz transition, and hence the critical field $E_c = E_c(c)$ for each concentration $c$ of the polymer. The results of the critical field of the Fréedericksz transition for both types of PSLC are shown in Figure 2. We normalized our data to the critical field $E_0 = \pi/d \sqrt{K/(\varepsilon_a \varepsilon_0)} = 0.24$ V/μm for BL038 in a planar geometry in the absence of a polymer network ($c=0$), where $\varepsilon_a = \varepsilon_\parallel - \varepsilon_\perp = 16.4$ is the dielectric anisotropy of the BL038 at 1 kHz, $\varepsilon_0$ is the permitivity of the free space, and $K$=20.7 pN is the Frank elastic constant for BL038 in the one constant approximation.

To explain these results we assume that the polymer network is composed of polymer fibrils oriented parallel to the nematic director in the field-off state. The concentration of the fibrils is proportional to the concentration of the dissolved diacrylate monomer. We also assume strong anchoring of the liquid crystal molecules on the surface of each fibril as well as on the surfaces of the glass substrates. By assuming that the polymer fibrils are perfectly straight we can treat the problem as two-dimensional. We model the domains of the nematic director in the field-on state by rectangles of dimension $a$ along the $x$-axis and $b$ along the $y$-axis. We consider the Fréedericksz transition in the PSLC cell of cell gap thickness $d$ in context of the transition in each of these identical rectangles defined by the director domains. Within each rectangular domain the total free energy density (per unit volume) of nematic liquid crystal can be written as:

$$f_v = \frac{K}{2}\left[\left(\frac{\partial \theta}{\partial x}\right)^2 + \left(\frac{\partial \theta}{\partial y}\right)^2 - \frac{\varepsilon_a \varepsilon_0 E^2}{K}\sin^2\theta\right] \quad (2)$$

where $\theta(x,y)$ is the distortion angle of the nematic director with respect to the $z$-axis. By minimizing the total free energy, we obtain the differential equation for the distortion angle:

$$\frac{\partial^2 \theta}{\partial x^2} + \frac{\partial^2 \theta}{\partial y^2} + \frac{\varepsilon_a \varepsilon_0 E^2}{K}\sin\theta\cos\theta = 0 \quad (3)$$

With the strong anchoring boundary conditions at the surface of the domain, i.e. $\theta$=0 at $x$=0, $a$ and $y$=0, $b$, the critical field $E_c$ is obtained by solving (3):

$$E_c = \sqrt{\frac{\pi^2 K}{\varepsilon_a \varepsilon_0}\left[\left(\frac{1}{a}\right)^2 + \left(\frac{1}{b}\right)^2\right]} \qquad (4)$$

We assume that the lengths $a$ and $b$ decrease with increasing concentration as: $1/a = 1/d + 1/\xi$ and $1/b = 1/\xi$, where the characteristic length $\xi$ is determined by the polymer network. The average distance between the centers of the randomly distributed polymer fibrils changes with the polymer concentration as $\alpha/\sqrt{c}$, where we express $c$ in %, and $\alpha$ will be determined from the data. Therefore, if we denote the diameter of the polymer fibrils by $\beta$ (see Figure 3(b)), the polymer characteristic length $\xi$ can be expressed as:

$$\xi = \frac{\alpha}{\sqrt{c}} - \beta \qquad (5)$$

With finite surface anchoring, which is the case on the perimeter of each of the rectangular director domains, the lengths $a$ and $b$ have to be modified. We assume that the surface anchoring energy per unit area takes the Rapini-Papoular form:[7]

$$f_s = 1/2\, W \sin^2 \theta \qquad (6)$$

where $W$ is the anchoring strength on the surfaces of the rectangular domains. The boundary conditions at finite anchoring are:[7,8]

$$\begin{aligned}\left(\frac{d\theta}{dx}\right)_a &\pm \frac{W}{K}\sin\theta\cos\theta = 0 \\ \left(\frac{d\theta}{dy}\right)_b &\pm \frac{W}{K}\sin\theta\cos\theta = 0\end{aligned} \qquad (7)$$

where the derivatives are taken on the perimeter of the rectangles. Hence the critical field with finite anchoring can be approximated as:

$$E_c = \sqrt{\frac{\pi^2 K}{\varepsilon_a \varepsilon_0} \left[ \left( \frac{1}{d} + \frac{1}{\xi + 2K/W} \right)^2 + \left( \frac{1}{\xi + 2K/W} \right)^2 \right]} \qquad (8)$$

This expression is valid when $K/W<10\xi$.[9] Comparing the elastic energy (2) and the surface energy (6) of the nematic liquid crystal on the boundaries of the nematic director domains where we assign the virtual surfaces: $f_v \beta \approx f_s$ (or equivalently $K(\theta/\xi)^2 \beta \approx W\theta^2$), we conclude that:

$$2\frac{K}{W} = A\frac{\xi^2}{\beta} \qquad (9)$$

where $A$ is a constant. At concentration $c=0$, the critical field in equation (8) gives the original Fréedericksz critical field $E_c = E_0 = \pi/d \sqrt{K/(\varepsilon_a \varepsilon_0)}$ as expected. By fitting the experimental data to the critical field of expression (8) we obtain $\alpha=0.21$ µm and $A=23.92$. The diameter $\beta=0.05$ µm of the polymer fibrils is estimated from the Scanning Electron Microscope pictures of PSLC.[3] The fit is shown in Figure 2 by the solid line. As our model assumes that the polymer fibrils do not touch each other, we can evaluate the polymer concentration above which our model is not valid by setting $\xi=0$. We find that this criterion corresponds to $c=17.72\%$. This concentration is already sufficiently high for the nematic liquid crystal to form a droplet like structure in the polymer matrix similar to Polymer Dispersed Liquid Crystals (PDLC),[1] where our model is not relevant. The anchoring strength value from our model $W = 1.95 \cdot 10^{-6} (1/\sqrt{c} - 0.24)^{-2} J/m^2$ would crossover to a concentration independent value in the PDLC case.

Another way to test the validity of our model is to measure the switching behavior of the PSLC. The time-off $\tau_{off}$ is related to the critical field $E_c$ defined in (8) by:

$$\tau_{off} E_c^2 = \frac{\pi^2 \gamma}{\varepsilon_a \varepsilon_0} \tag{10}$$

where $\gamma$ is the rotational viscosity of the liquid crystal. For each concentration the time-off was measured after application of the field of strength corresponding to the voltage $V^*$ (see Figure 1(b)), where the light transmittance $T$ reaches its minimum for the first time, to insure that for each concentration of the polymer the phase retardation is the same at the maximum applied voltage $V^*$. Figure 4 shows the experimental results as well as the fit corresponding to (10). From the fit we calculate the rotational viscosity of BL038 to be 0.33 Pa·s. The time-off on the order of 3 ms is achievable for 6 % polymer concentration. The comparable measured values of the critical field and the time-off for the two types of PSLC studied (LC242 and RM257) suggest the existence of a common polymer network morphology and influence on the nematic host.

This work was supported by the National Science Foundation under grants DMR-9875427 (G. P. C and P. A. K.) and DMR-9873849 (N. V. P. and R. A. P.).

**References**

1. Liquid Crystals in Complex Geometries, Edited by G. P. Crawford and S. Zumer (Taylor & Francis, London, 1996)

2. C.-C. Chang, L.-C. Chien, and R. B. Meyer, Phys. Rev. E **56**(1), 595-599 (1997)

3. R.-Q. Ma, D.-K. Yang, Phys. Rev. E **61**(2), 1567-1573 (2000)

4. M. J. Escuti, C. C. Bowley, G. P. Crawford, S. Zumer, Appl. Phys. Lett. **75**(21), 3264-3266 (1999)

5. R. A. M. Hikmet, H. M. J. Boots, Phys. Rev. E **51**(6), 5824-5831 (1995)

6. R. A. M. Hikmet, J. Appl. Phys. **68**(9), 4406-4412 (1990)

7. A. Rapini, M. J. Papoular, J. Phys. (Paris), Colloq. C4, **30**, C4-54 (1969)

8. J. Nehring, A. R. Kmetz, T. J. Scheffer, J. Appl. Phys. **47**, 850 (1976)

9. L. M. Blinov, V. G. Chigrinov, Electrooptic Effects in Liquid Crystal Materials, Chapter 3, 111-113, Springer 1993

**Captions**

Figure 1:

(a) The experimental set up for measuring the transmission properties of a PSLC cell. The cell is situated between cross polarizers oriented at 45º to the nematic director *n* inside the cell. (b) An example of the transmittance versus applied voltage scans.

Figure 2:

Critical field versus concentration for two PSLC prepared using BL038 and diacrylate monomers LC242 (BASF), and RM257 (EM Industries). $E_c$ is the critical field normalized to the critical field $E_0$ in the planar cell filled with pure BL038. The solid curve represents the fitting based on the equation (8).

Figure 3:

(a) The polymer network is shown as a collection of fibrils parallel to the nematic director *n* in the field-off state (*z*-axis). The electric field *E* was applied along the *x*-axis perpendicular to the glass substrates. The director domains are modeled by parallelepipeds with rectangular cross-section of dimensions *a* along the *x*-axis and *b* along the *y*-axis. (b) Average distance between the centers of polymer fibrils at concentration *c* is $\alpha/\sqrt{c}$, and the diameter of a fibril is $\beta$.

Figure 4:

The time-off results for both types of PSLC as a function of polymer concentration. The solid curve represents the fitting based on the equation (10).

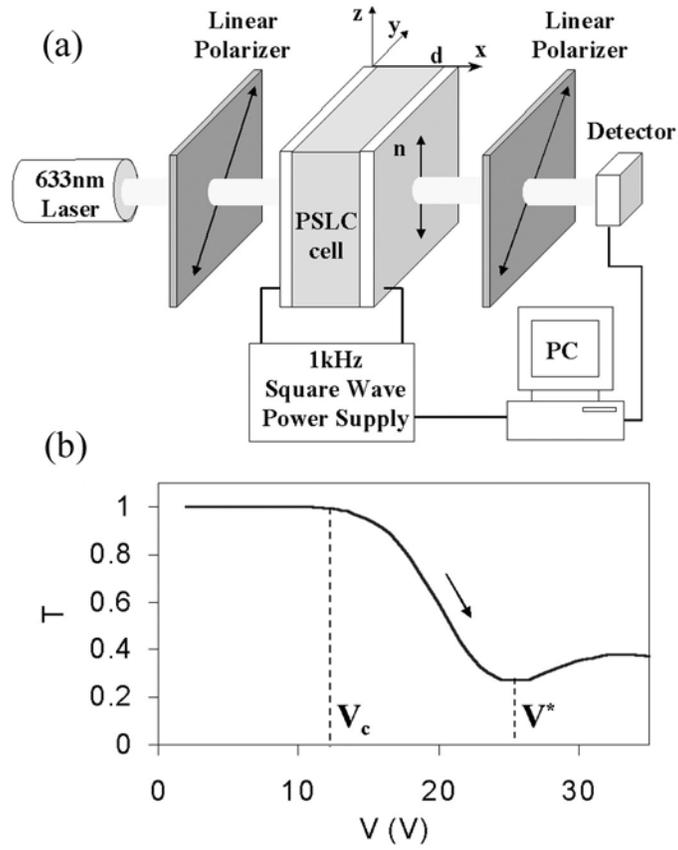

Figure 1 (Kossyrev *et al.*)

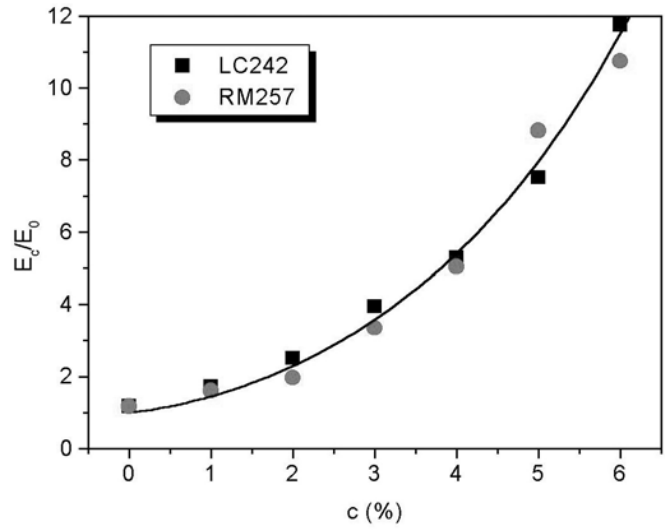

Figure 2 (Kossyrev *et al*.)

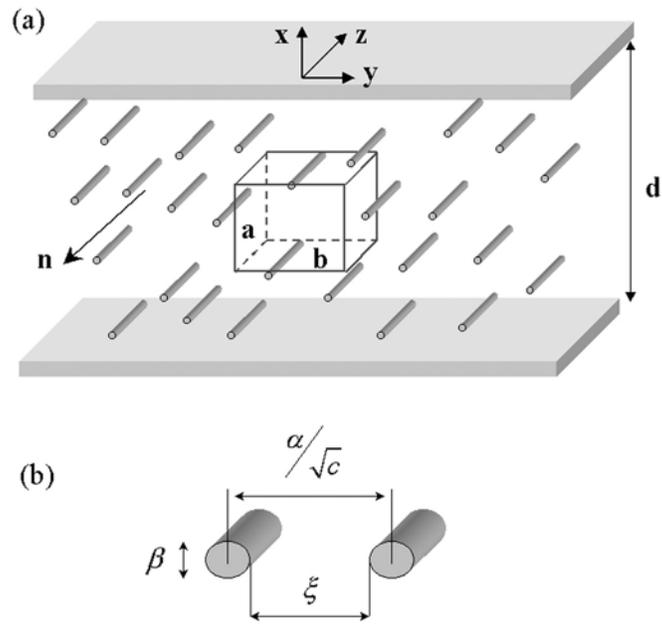

Figure 3 (Kossyrev *et al.*)

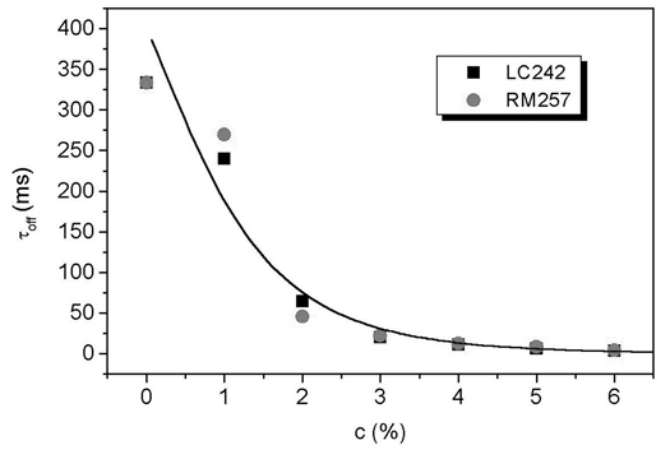

Figure 4 (Kossyrev *et al.*)